\documentclass[runningheads]{llncs}
\usepackage[T1]{fontenc}
\usepackage{multirow}
\usepackage{graphicx}
\usepackage[pagebackref=true,breaklinks=true,colorlinks,bookmarks=false]{hyperref}
\begin{document}
\title{Iterative Semi-Supervised Learning for Abdominal Organs and Tumor Segmentation}
%
%
\author{Jiaxin Zhuang\inst{1}\orcidID{0000-0001-9287-4263} \and
Luyang Luo\inst{1}\orcidID{0000-0002-7485-4151} \and
Zhixuan Chen\inst{1}\orcidID{0000-0001-8767-7177}, Linshan Wu\inst{1}\orcidID{0000-0002-0486-184X} 
} 
\authorrunning{Jiaxin Zhuang et al.}
%
\institute{The Hong Kong University of Science and Technology, Hong kong, China
\email{jzhuangad@connect.ust.hk}}

\maketitle              
\begin{abstract}
Deep-learning (DL) based methods are playing an important role in the task of abdominal organs and tumors segmentation in CT scans. However, the large requirements of annotated datasets heavily limit its development. The FLARE23 challenge provides a large-scale dataset with both partially and fully annotated data, which also focuses on both segmentation accuracy and computational efficiency. In this study, we propose to use the strategy of Semi-Supervised Learning (SSL) and iterative pseudo labeling to address FLARE23. Initially, a deep model (nn-UNet) trained on datasets with complete organ annotations (about 220 scans) generates pseudo labels for the whole dataset. These pseudo labels are then employed to train a more powerful segmentation model. Employing the FLARE23 dataset, our approach achieves an average DSC score of 89.63\% for organs and 46.07\% for tumors on online validation leaderboard. For organ segmentation, We obtain 0.9007\% DSC and 0.9493\% NSD. For tumor segmentation, we obtain 0.3785\% DSC and 0.2842\% NSD. Our code is available at \href{https://github.com/USTguy/Flare23}{here}.

\end{abstract}

\section{Introduction}
The tumor growth in abdomen has received significant attention recently. Deep learning (DL)
 based methods have achieved promising ability to tumor and organ segmentation. However, adequate and accurate annotation of tumors and relevant abdominal organs in CT scans are still very expensive which heavily hinders the performance of DL. Specifically, there are several challenging problems in this field. First, the lack of datasets that include annotations for both tumors and abdominal organs, \emph{i.e.}, existing datasets mainly contain only organ or tumor anntoations. Thus, it is difficult to learn a robust segmentation model from only partially labeled and unlabeled datasets. Second, although the state-of-the-art solution, \emph{i.e.},nnU-Net has demonstrated promising results, it is still very time-consuming, which heavily limits its practical utility. To address these problems, the FLARE23 challenge (Fast, Low-resource, and Accurate oRgan and Pan-cancer sEgmentation in Abdomen CT) has been established, which provides a large-scale dataset that includes both partially annotated and unlabeled data.

\begin{figure}[htbp]
	\centering
	\includegraphics[width=1\linewidth]{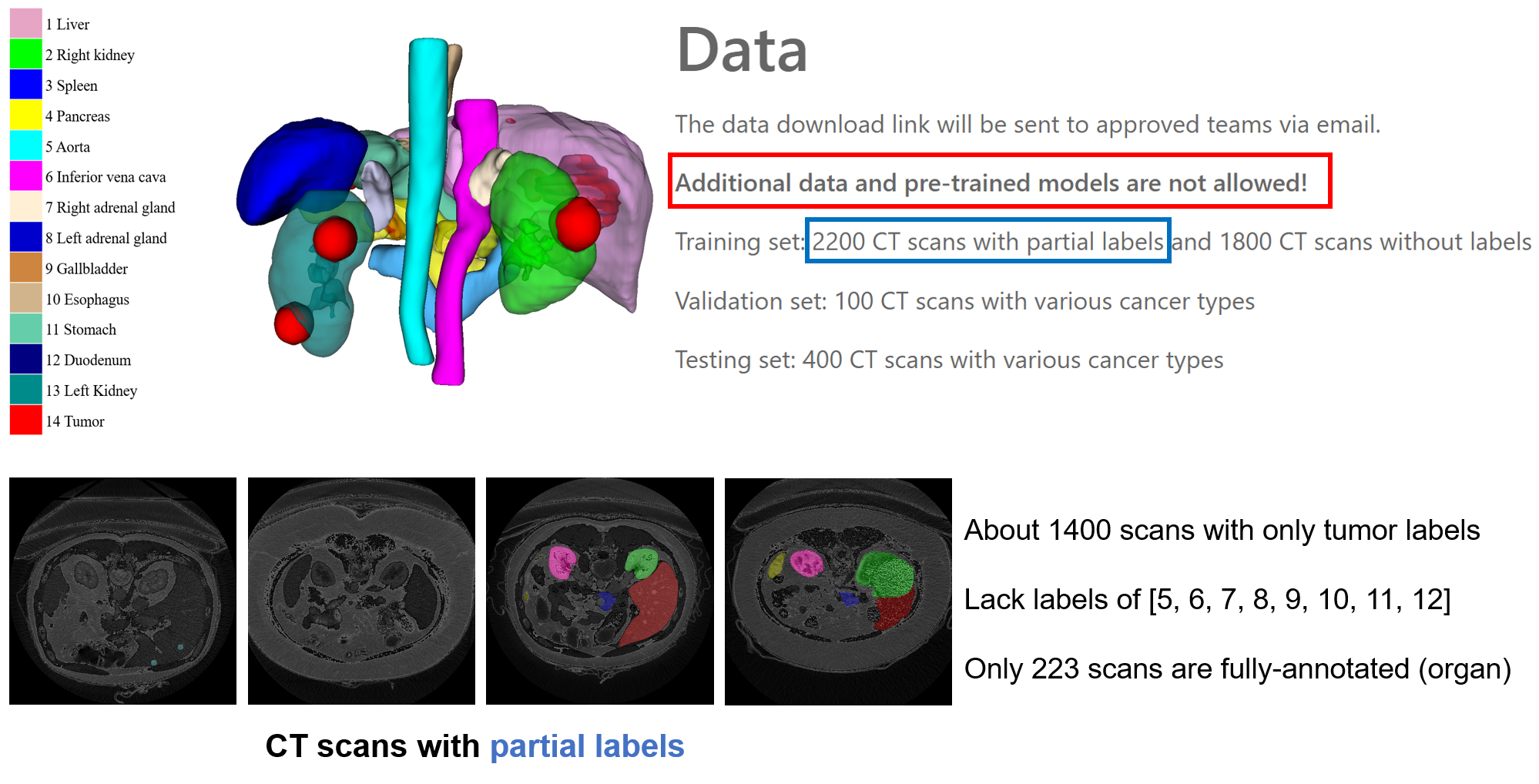}
	\caption{\textbf{The brief information of FLARE23}. FLARE23 dataset includes 4000 3D CT scans from 30+ medical centers. 2200 cases have partial labels and 1800 cases are unlabeled. }
	\label{fig_info}
\end{figure}

FLARE23 (Fast, Low-resource, and Accurate oRgan and Pan-cancer sEgmentation in Abdomen CT) aims to promote the development of universal organ and tumor segmentation in abdominal CT scans. The brief information of FLARE23 is shown in Fig.~\ref{fig_info}. Specifically, the most challenging problem is the partially labeled scans. In 2200 labeled scans of FLARE23, about 1400 cases contain only tumor annotations. Only 223 cases are fully-annotated with organs. Thus, the most essential issue is to solve the problem of few and partial labels.

In our contest, we try to use solve the problems with Semi-Supervised Learning (SSL)~\cite{SSL1,SSL2,SSL3,wu2022deep} with iterative pseudo labels refinement~\cite{SSL1,SSL2,wu2023sparsely,wu2022deep}, which is a typical solution for dataset with only limited labeled samples. SSL first trains a teacher model with only labeled data then employ the teacher model to generate pseudo labels for the unlabeled data. Finally, SSL further train a student model on both labeled data and pseudo-labeled data, which can obtain a more powerful model. However, there still exists a challenging problem. With only few accurate labeled data for training a teacher model, we cannot effectively guarantee the quality of generated pseudo labels of unlabeled data. Thus, it is important to figure out a more effective way to generate pseudo labels.

In our contest, we propose to use an iterative SSL framework to refine the generated pseudo labels step by step, which is a multi-stage process. We observe that the performance is increasing consistently when we rectify the pseudo labels using iterative training. However, we still cannot achieve promising results finally. And our proposed method also requires very time-consuming training process, which is not efficient for practical application. Thus, we fail to submit a good result in the final stage. The details of our proposed method are presented in Section~\ref{sec3}.

\section{Method}
\label{sec3}
\begin{figure}[htbp]
\centering
\includegraphics[scale=0.22]{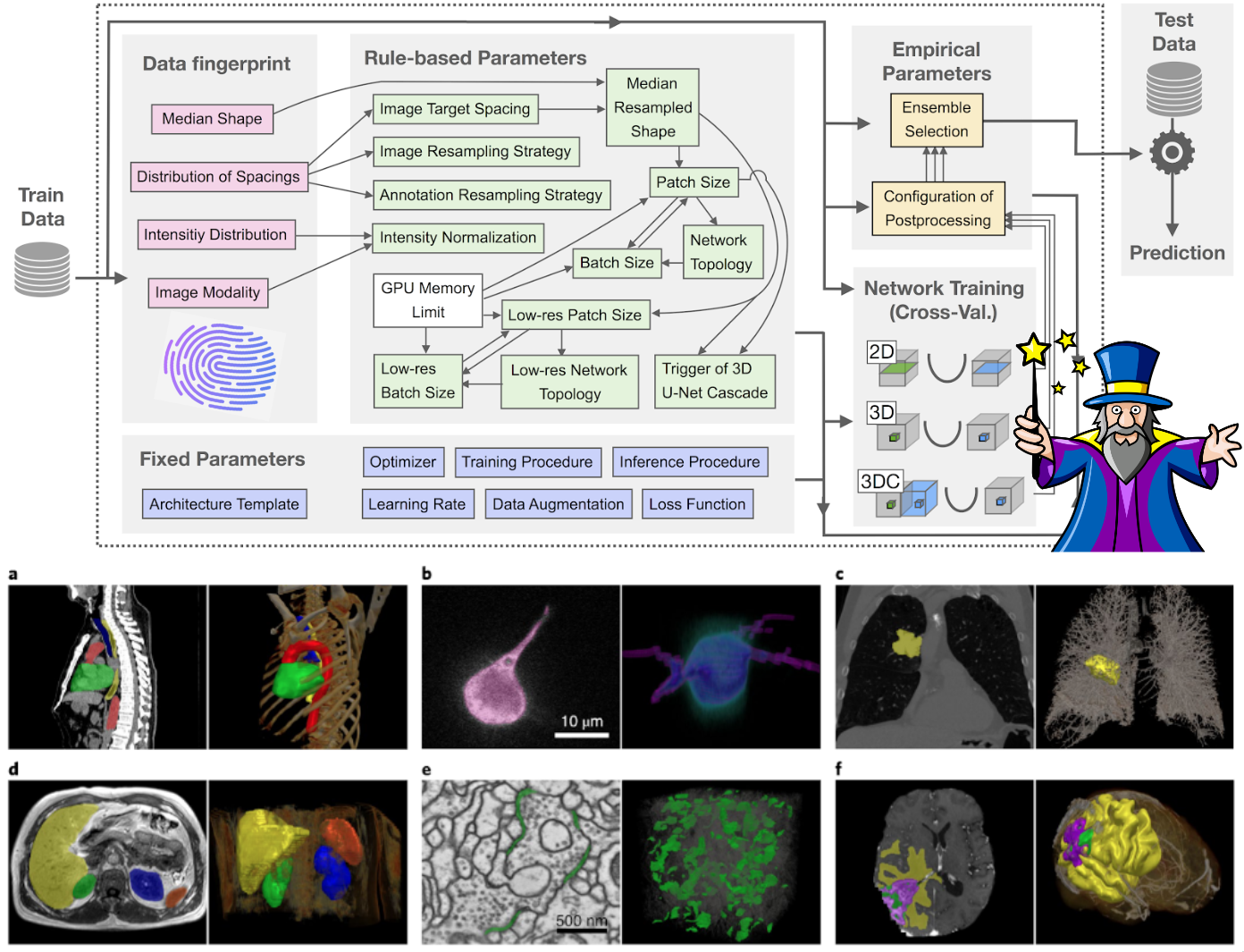}
\caption{The pipeline of nnU-Net~\cite{nnUNet}, including pre-processing, training, inference, and post-processing.
}
\label{fig:nnunet}
\end{figure}

Following previous execellent solutions~\cite{FLARE22-1st-Huang,FLARE22-bestDSC-Wang} in Flare21 and Flare22, we also use nnU-Net~\cite{nnUNet} in our contest. Figure~\ref{fig:nnunet} shows a typical example of 3D nnU-Net~\cite{nnUNet}. We also use the default pre-processing and post-processing methods of 3D nnU-Net~\cite{nnUNet}. The details are as follows.

\subsection{Preprocessing}
We also use the default preprocessing of nnU-Net~\cite{nnUNet}. Specifically, for anisotropic data resampling, trilinear interpolation is used in the axial plane and linear interpolation in the sagittal direction. Intensity normalization is performed by clipping values to the 0.5\% (-970.0) and 99.5\% (279.0) Hounsfield Unit levels, followed by z-normalization using a mean of 80.3 and a standard deviation of 141.4.

\subsection{Proposed Method}
\begin{figure}[htbp]
\centering
\includegraphics[scale=0.35]{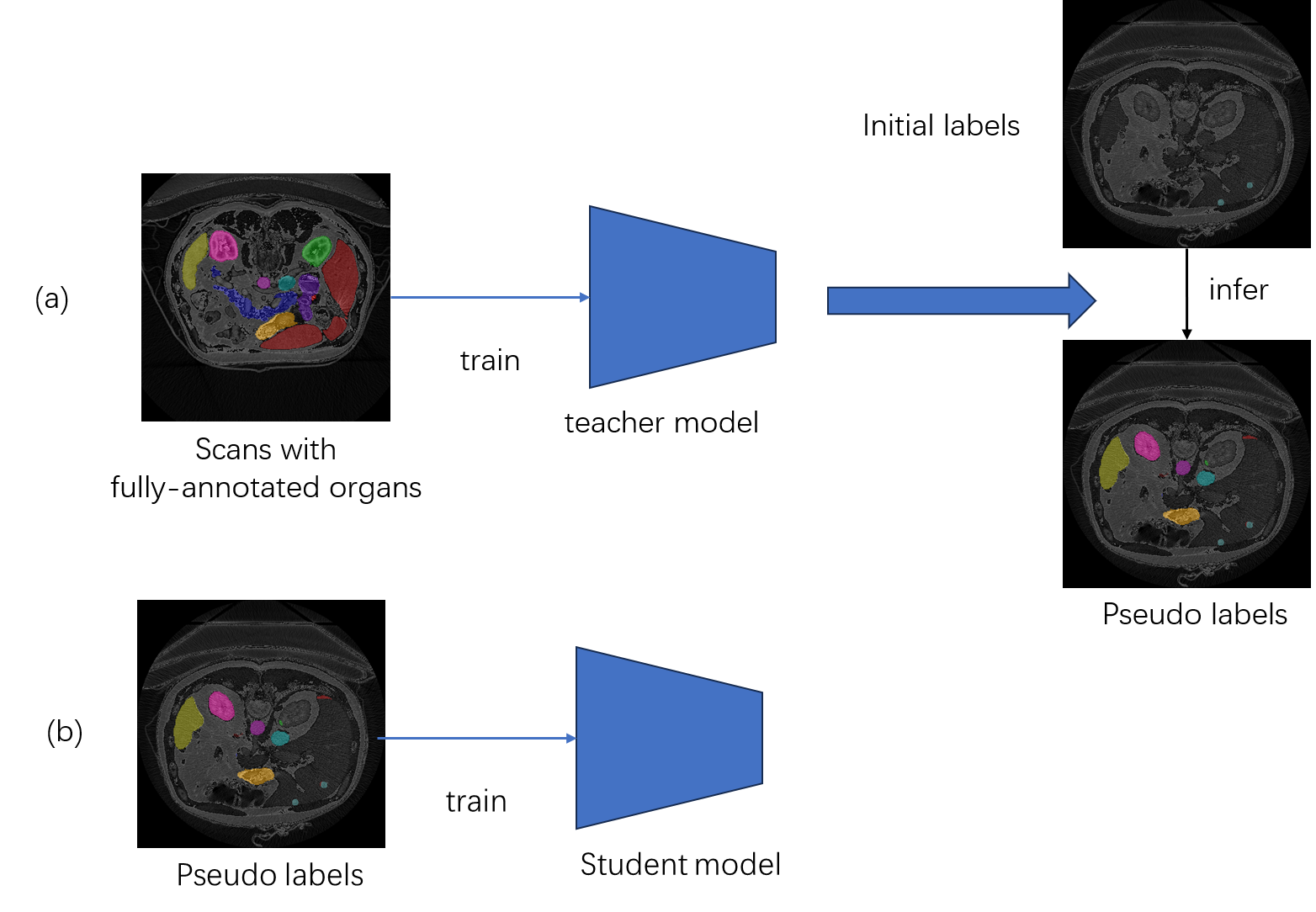}
\caption{Network architecture (Copyright preserved. Please do not directly use this figure in your manuscript.) Please also include the network description in the figure title. So reviewers could quickly understand your idea. 
}
\label{fig:frame}
\end{figure}

Inspired by the winning solution of FLARE 2022, we implement a multi-stage framework to generate pseudo-labels for unlabeled data. Specifically, as described in Fig.~\ref{fig_info}, there are only 223 scans with fully-annotated organs and about 1400 scans with only tumor annotations. Thus, our proposed iterative SSL method contains two phases: one phase to generate organ pseudo labels while the other will generate tumor pseudo labels. The details are illustrated in Fig.~\ref{fig:frame}.

As shown in Figs.~\ref{fig:frame}, our framework contains two parts. First, we train a model on 1400 scans with tumor annotations. Then, this model will generate pseudo tumor labels for other images. Then, we will use all images to train a new student model for tumor segmentation. This process works in an iterative manner. Similarity, we employ this process to train a model for organ segmentation. Finally, we ensemble the tumor and organ annotations together for final training.

We also use the pseudo labels generated by the FLARE21 winning algorithm~\cite{FLARE22-1st-Huang} and the best-accuracy-algorithm~\cite{FLARE22-bestDSC-Wang}. Specifically, we ensemble these pseudo labels and our own generated pseudo labels together, aiming to obtain more accurate and reliable supervision for the unlabeled data.

For Loss function, we use the standard loss in nnU-Net~\cite{nnUNet}: summation between Dice loss and cross-entropy loss because compound loss functions have been proven to be robust in various medical image segmentation tasks~\cite{LossOdyssey}. 

\textbf{Limitation:} we do not figure out how to deal with the partial labels well. We simply ensemble them with our pseudo labels. But the final performances are not promising.

\textbf{Limitation:} we do not develop good strategies to improve inference speed and reduce resource consumption. We still follow the inference procedure of nnU-Net~\cite{nnUNet} in the validation and testing.

\subsection{Post-processing}
We also use the default post-processing of nnU-Net~\cite{nnUNet}. During the pseudo-labeling generation phase, we employed Testing Time Augmentation (TTA) along the anatomical axes: sagittal, coronal, and axial, to enhance the quality of the generated labels.

\section{Experiments}
\subsection{Dataset and evaluation measures}
The FLARE 2023 challenge is an extension of the FLARE 2021-2022~\cite{MedIA-FLARE21}\cite{FLARE22}, aiming to aim to promote the development of foundation models in abdominal disease analysis. The segmentation targets cover 13 organs and various abdominal lesions. The training dataset is curated from more than 30 medical centers under the license permission, including TCIA~\cite{TCIA}, LiTS~\cite{LiTS}, MSD~\cite{simpson2019MSD}, KiTS~\cite{KiTS,KiTSDataset}, and AbdomenCT-1K~\cite{AbdomenCT-1K}. The training set includes 4000 abdomen CT scans where 2200 CT scans with partial labels and 1800 CT scans without labels. The validation and testing sets include 100 and 400 CT scans, respectively, which cover various abdominal cancer types, such as liver cancer, kidney cancer, pancreas cancer, colon cancer, gastric cancer, and so on. The organ annotation process used ITK-SNAP~\cite{ITKSNAP}, nnU-Net~\cite{nnUNet}, and MedSAM~\cite{MedSAM}.

The evaluation metrics encompass two accuracy measures—Dice Similarity Coefficient (DSC) and Normalized Surface Dice (NSD)—alongside two efficiency measures—running time and area under the GPU memory-time curve. These metrics collectively contribute to the ranking computation. Furthermore, the running time and GPU memory consumption are considered within tolerances of 15 seconds and 4 GB, respectively.

\subsection{Implementation details}
\subsubsection{Environment settings}
The development environments and requirements are presented in Table~\ref{table:env}.

\begin{table}[!htbp]
\caption{Development environments and requirements.}\label{table:env}
\centering
\begin{tabular}{ll}
\hline
System       & Ubuntu 18.04.5 LTS\\
\hline
CPU   & Intel(R) Core(TM) i9-7900X CPU@3.30GHz \\
\hline
RAM &16$\times $4GB; 2.67MT$/$s\\
\hline
GPU (number and type) & one NVIDIA 3090Ti 24G\\
\hline
CUDA version & 11.0\\ 
\hline
Programming language & Python 3.20\\ 
\hline
Deep learning framework & torch 2.0, torchvision 0.2.2 \\ 
\hline
Code     &https://github.com/USTguy/Flare23\\
\hline
\end{tabular}
\end{table}

\subsubsection{Training protocols}
The training protocols are mainly inherited from nnU-Net~\cite{nnUNet}. Specifically, we utilize the preprocessing and pseudo-labeling scheme discussed earlier. In addition, we adopt extensive data augmentation techniques, including rotations, elastic deformations, and random cropping, to enhance our models’ generalization capabilities.

\begin{table*}[!htbp]
\caption{Training protocols.}
\label{table:training}
\begin{center}
\begin{tabular}{ll} 
\hline
Network initialization &He \\
\hline
Batch size &2 \\
\hline 
Patch size &48$\times$192$\times$192  \\ 
\hline
Total epochs & 1000\\
\hline
Optimizer  &SGD\\ 
\hline
Initial learning rate (lr)  &0.01\\ 
\hline
Lr decay schedule &poly decay\\
\hline
Training time  & 120 hours \\  
\hline 
Loss function & DiceCEloss \\
\hline
Number of model parameters    & 440M\footnote{https://github.com/sksq96/pytorch-summary} \\ \hline
Number of flops & 3.81T\footnote{https://github.com/facebookresearch/fvcore} \\ \hline
CO$_2$eq & 114.02Kg\footnote{https://github.com/lfwa/carbontracker/} \\  \hline
\end{tabular}
\end{center}
\end{table*}

\section{Results and discussion}

\begin{table}[htbp]
\caption{Quantitative evaluation results. \textbf{The public validation denotes the performance on the 50 validation cases with ground truth. Please present both the mean score and standard deviation. The online validation denotes the leaderboard results. The Testing results will be released during MICCAI. Please leave them blank at present.} You can use a similar 
Table format to present the ablation study results of the public and online validation. A useful online tool to create latex table \url{https://www.tablesgenerator.com/latex_tables.}
}
\centering
\begin{tabular}{l|cc|cc|cc}
\hline
\multirow{2}{*}{Target} & \multicolumn{2}{c|}{Public Validation} & \multicolumn{2}{c|}{Online Validation} & \multicolumn{2}{c}{Testing} \\ \cline{2-7} 
& DSC(\%)            & NSD(\%)           & DSC(\%)            & NSD(\%)           & DSC(\%)      & NSD (\%)     \\ \hline
Liver & 0.9694  &0.9787                  &                  &                &            &              \\
Right Kidney &0.9484 &0.9550                   &                    &                   &              &              \\
Spleen &0.9788 &0.9928                   &                    &                   &              &              \\
Pancreas &0.8534 &0.9595                   &                    &                   &              &              \\
Aorta &0.9591 &0.9854                   &                    &                   &              &              \\
Inferior vena cava &0.9319                    & 0.9505                  &                    &                   &              &              \\
Right adrenal gland     &0.8866 &0.9672                   &                    &                   &              &              \\
Left adrenal gland      &0.8761 &0.9522                   &                    &                   &              &              \\
Gallbladder             &0.8587                    &0.8600                   &                    &                   &              &              \\
Esophagus               &0.7831                    &0.9091                   &                    &                   &              &              \\
Stomach                 &0.9158                    &0.9478                   &                    &                   &              &              \\
Duodenum                &0.8051                    &0.9338                   &                    &                   &              &              \\
Left kidney             &0.9429                    &0.9484                   &                    &                   &              &              \\
Tumor                   &0.3785                    &0.2842                   &                    &                   &              &              \\ \hline
Organ-Average                   &0.9007                    &0.9493                   &                    &                   &              &              \\ \hline
\end{tabular}
\label{result}
\end{table}

\subsection{Quantitative results on validation set}
We report the Dice and NSD scores of organs and tumors on the validation set, as shown in Table~\ref{result}. For organ segmentation, We obtain 0.9007\% DSC and 0.9493\% NSD. For tumor segmentation, we obtain 0.3785\% DSC and 0.2842\% NSD. It can be seen that the results of tumor segmentation are not so good, which heavily limits our final performance.

\subsection{Qualitative results}

\begin{figure}[htbp]
\centering
\includegraphics[scale=0.35]{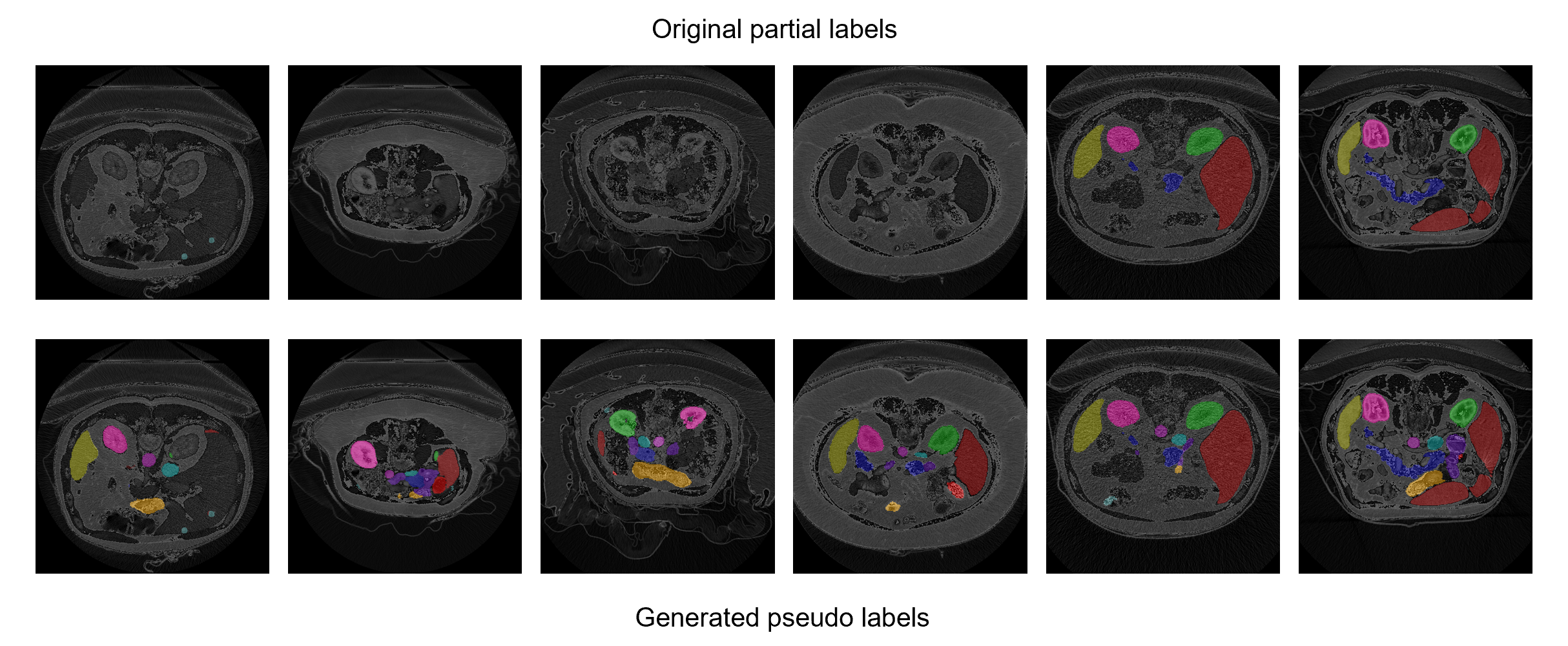}
\caption{Visualization results of our generated pseudo labels, comparing with the initial labels.
}
\label{fig:pseudo}
\end{figure}

The visualization results of our proposed iterative SSL in Flare23 dataset are shown in Fig.~\ref{fig:pseudo}. As can be seen, with our iterative SSL method, we can generate more complete and accurate pseudo labels, which can provide stronger supervision than the original partial labels.

\subsection{Segmentation efficiency results on validation set}

We did not evaluate the efficiency results on validation set. Since we use the settings of nn-UNet, the efficiency is very bad. We will explore it in the future.

\subsection{Limitation and future work}

Obviously, we fail in this Flare23 contest. The accuracy and efficiency are both not so good. The generated pseudo labels of tumor segmentation are not reliable enough for the learning of unlabeled data, which heavily limits the accuracy of tumor segmentation. We have learned a lot from others' reports. We will try more effective SSL methods in the future contests. And we will also try to boost the efficiency by improving the inference strategies in the future.

\section{Conclusion}
In this contest, we find that iterative SSL can significantly improve the performance. And the ensemble of pseudo labels can also gain obvious improvements. Although we fail the contest, Flare23 still provides us a lot of valuable experience.

\subsubsection{Acknowledgements} The authors of this paper declare that the segmentation method they implemented for participation in the FLARE 2023 challenge has not used any pre-trained models nor additional datasets other than those provided by the organizers. The proposed solution is fully automatic without any manual intervention. We thank all the data owners for making the CT scans publicly available and CodaLab~\cite{codalab} for hosting the challenge platform.

%
%
%
\bibliographystyle{splncs04}
\bibliography{ref}

\newpage
\begin{table}[!htbp]
\caption{Checklist Table. Please fill out this checklist table in the answer column.}
\centering
\begin{tabular}{ll}
\hline
Requirements                                                                                                                    & Answer        \\ \hline
A meaningful title                                                                                                              & Yes        \\ \hline
The number of authors ($\leq$6)                                                                                                             & 4        \\ \hline
Author affiliations, Email, and ORCID                                                                                           & Yes        \\ \hline
Corresponding author is marked                                                                                                  & Yes        \\ \hline
Validation scores are presented in the abstract                                                                                 & Yes        \\ \hline
\begin{tabular}[c]{@{}l@{}}Introduction includes at least three parts: \\ background, related work, and motivation\end{tabular} & Yes        \\ \hline
A pipeline/network figure is provided                                                                                           & 4 \\ \hline
Pre-processing                                                                                                                  & 3   \\ \hline
Strategies to use the partial label                                                                                             & 4   \\ \hline
Strategies to use the unlabeled images.                                                                                         & 4   \\ \hline
Strategies to improve model inference                                                                                           & 5   \\ \hline
Post-processing                                                                                                                 & 5   \\ \hline
Dataset and evaluation metric section is presented                                                                              & 5   \\ \hline
Environment setting table is provided                                                                                           & 6  \\ \hline
Training protocol table is provided                                                                                             & 5  \\ \hline
Ablation study                                                                                                                  & 6   \\ \hline
Visualized segmentaiton example is provided                                                                                     & 8 \\ \hline
Limitation and future work are presented                                                                                        & Yes        \\ \hline
Reference format is consistent.  & Yes        \\ \hline

\end{tabular}
\end{table}

\end{document}